\documentclass[amsmath,amssymb,aps,prb,twocolumn,superscriptaddress,10pt]{revtex4-2}

\usepackage{graphicx}
\usepackage{wasysym}
\usepackage{color}
\usepackage[normalem]{ulem}

\begin{document}

\title{Topological transitions in spin-ice induced by geometrical constraints}
	 
\author{R. A. Borzi}
\affiliation{Instituto de Física de Líquidos y Sistemas Biológicos (IFLYSIB), UNLP-CONICET, Argentina.} 
\affiliation{Departamento de Física, Facultad de Ciencias Exactas, Universidad Nacional de La Plata, Argentina.}
\author{E. S. Loscar }
\affiliation{Instituto de Física de Líquidos y Sistemas Biológicos (IFLYSIB), UNLP-CONICET, Argentina.} 
\affiliation{Departamento de Física, Facultad de Ciencias Exactas, Universidad Nacional de La Plata, Argentina.}

\author{S. A. Grigera}
\affiliation{Instituto de Física de Líquidos y Sistemas Biológicos (IFLYSIB), UNLP-CONICET, Argentina.} 
\affiliation{Departamento de Física, Facultad de Ciencias Exactas, Universidad Nacional de La Plata, Argentina.}
	
\date{\today}

\begin{abstract}
We study the nearest-neighbor spin-ice model subjected to a magnetic field applied along the global [111] and [110] directions, focusing on the role of sample geometry in stabilizing topological phase transitions. While no Kasteleyn transition is expected for this field orientations in the thermodynamic limit, we show that constraining the transverse dimensions of the system qualitatively changes the behavior. For samples elongated along the field direction with finite transverse area, the divergence-free constraint quantizes the number of string excitations that can span the system. As a result, the magnetization evolves through a cascade of discrete transitions corresponding to the successive entry of individual strings. Using Monte Carlo simulations, we demonstrate that each transition is marked by sharp magnetization steps and peaks in the specific heat and susceptibility, whose amplitudes scale linearly with the system length. We complement the numerical results with an analytical treatment based on the entropy–energy balance on a system with reduced dimensionality,
deriving the critical fields associated with each topological sector. In the isotropic limit these transitions merge into a smooth crossover, but for anisotropic samples they remain sharply resolved, illustrating an unconventional mechanism by which finite geometry stabilizes topological phase transitions in frustrated magnets.
\end{abstract}
	
\maketitle

\section{Introduction}

Magnetic systems provide a fertile setting for the study of statistical mechanics models, offering realizations of a wide range of phenomena such as glassy dynamics~\cite{kawashima2013recent}, Bose–Einstein condensation~\cite{zapf2014bose}, emergent gauge fields~\cite{moessner2021spin}, spin liquids~\cite{balents2010spin,knolle2019field,broholm2020quantum} magnetic monopole liquids~\cite{castelnovo2012spin,slobinsky2018charge,slobinsky2021monopole}, and Kasteleyn transitions in two and three dimensions~\cite{moessner2003theory,isakov2004magnetization,jaubert2008three,szabo2025hidden}.

Kasteleyn transitions were first identified in dimer lattices~\cite{kasteleyn1963dimer} and later rediscovered in other systems~\cite{nagle1973lipid}. The divergence-free constraint of the spin-ice ground state provided a natural route to realizing these transitions in frustrated magnets. A two-dimensional Kasteleyn transition was first found for fields slightly misaligned from the [111] direction~\cite{moessner2003theory,fennell2007pinch}, while a three-dimensional analogue was proposed and observed for fields along [100]~\cite{jaubert2008three,Morris2009,pili2022topological}. Subsequent studies extended these ideas to general field orientations, showing that Kasteleyn-like transitions occur for a broad range of directions \cite{benton2022spin}. In spite of this, and the obvious fact that the divergenceless condition should be valid at moderate fields for any field direction, it is well known that no such transitions are found in the three dimensional thermodynamic limit for the high symmetry directions [111] and [110]~\cite{fukazawa2002magnetic,hiroi2003ferromagnetic,moessner2003theory,melko2004monte,sakakibara2021magnetic,benton2022spin}.

Here we focus on the nearest-neighbor spin-ice model in the presence of an external magnetic field applied along the global [111] direction. This geometry has been extensively studied~\cite{moessner2003theory,isakov2004magnetization,Sakakibara2003,saito2005magnetodielectric,sato2007field,kao2016field,Borzi2016,vignau2023numerical}, as the interplay between local ice-rule constraints and the external field produces a variety of ordered and disordered phases. In particular, the field selects the kagome-ice manifold at intermediate strengths, while stronger fields drive the system into a fully polarized state. As noted above, no topological phase transitions are found for this field orientation in the thermodynamic limit. Conventional wisdom—familiar from textbook examples such as Bose–Einstein condensation or Ising ferromagnetism—holds that finite-size geometric constraints tend to suppress or smear phase transitions, including those of Kasteleyn type~\cite{jaubert2008three,baez20163d,szabo2025hidden}. Here, however, we find the opposite behavior. At low fields, the finite geometry together with a global constraint stabilizes a genuine topological phase transition or, more precisely, a cascade of discrete transitions. This mechanism reveals a qualitatively new route to the emergence of topology in frustrated materials. We further show that an analogous effect arises when the external field is applied along the [110] direction.

The key idea is that the entropy associated with the creation of each segment of a topological string excitation running parallel to the field is stored in the direction perpendicular to it. Unlike the \([100]\) case, where each link in the chain contributes only a fixed entropy \(\log(2)\)~\cite{jaubert2008three}, this entropy grows without bound for \(\mathbf{B}\!\parallel\![111]\) or \([110]\) when the thermodynamic limit is taken in all three spatial dimensions. Confining the crystal in the perpendicular direction fundamentally alters this situation, yielding a topological first order phase transition at finite values of the single thermodynamic control parameter, namely the ratio of field to temperature. In this sense, the two field orientations we study are complementary. For \(\mathbf{B}\!\parallel\![110]\), the crucial constraint acts along the direction of one-dimensional chains, while for \(\mathbf{B}\!\parallel\![111]\) the entropy gain is associated with Kagome planes.

Given the entropic repulsion between chains, a second important consequence of the geometrical constriction emerges. A finite transverse dimension quantizes the number of topological string excitations that can span the system. As the field or temperature is varied, strings enter the sample one by one, producing a sequence of sharp magnetization steps and peaks in the specific heat, magnetic susceptibility and density of monopoles, as previously observed in two-dimensional spin ice~\cite{ferreyra2018breakdown}. Each step marks the limit of a distinct topological sector, separated by an energy gap that scales linearly with the system length along the field direction. In the isotropic thermodynamic limit these discrete transitions merge into a smooth crossover, but for samples with reduced dimensionality, they can remain individually resolvable. This finite-size stabilization of a topological cascade contrasts with the more common rounding of transitions by finite-size effects and provides a clear example of geometry-driven topology in a frustrated magnet.

%%%%%%%%%%%%%%%%%%%    
\begin{figure}[b!t]
\centering
\includegraphics[width=0.6\linewidth]{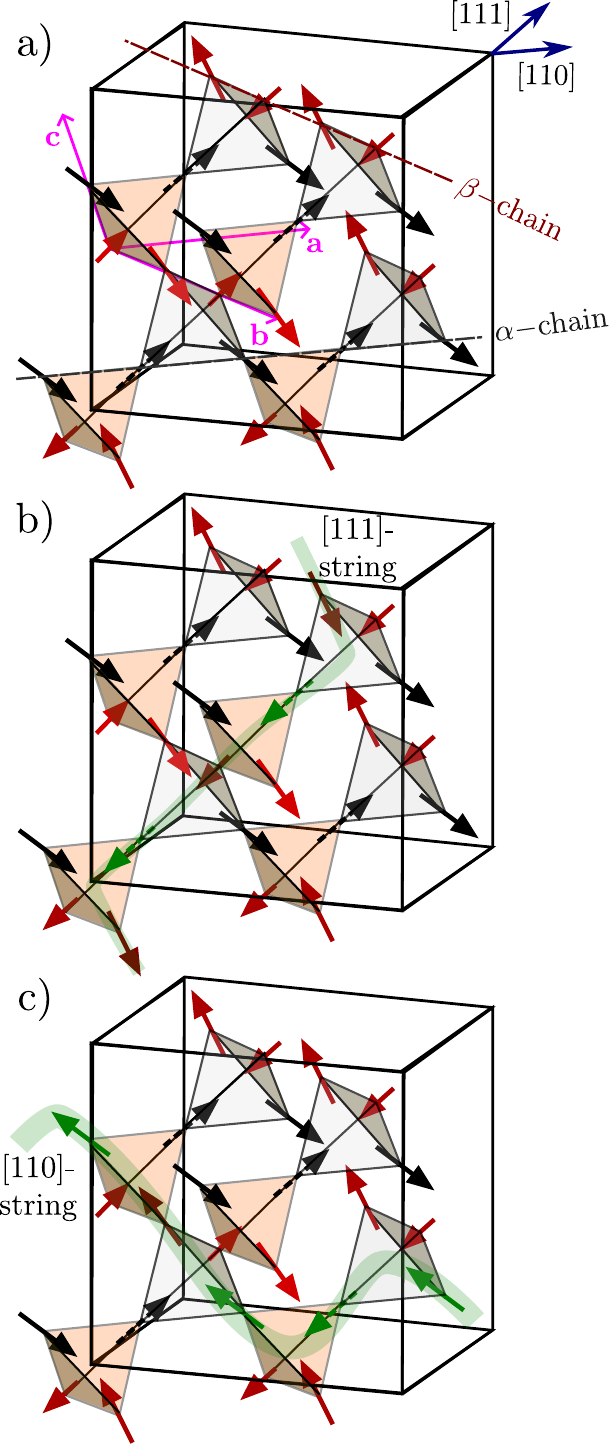}
\caption{\textit{a-} A conventional, cubic unit cell for the pyrochlore lattice, made of one set of tetrahedra pointing down (marked in gray) and another one pointing up. The two-in/two-out spin configuration is compatible with a saturated state both for $\mathbf{B}\parallel[111]$ or $\mathbf{B}\parallel[110]$. For $\mathbf{B}\parallel[111]$ the lattice can be thought as made of spins on triangular planes (dashed arrows) and spins on Kagome planes (dark triangles). For $\mathbf{B}\parallel[110]$, it can be thought as composed of $\alpha-$chains (lines of black spins, with a component in the direction of [110]) and $\beta-$chains (lines of red spins perpendicular to the field).  The cell constants $\mathbf{a}$, $\mathbf{b}$, and $\mathbf{c}$ are indicated in magenta. They allow for periodic boundary conditions connecting the sides of the Kagome planes and the direction along $\mathbf{a}\parallel[110]$. \textit{b-} A first [111]-string excitation enters the system, changing the topology of the system in a phase transition; it consists of a succession of steps involving inverting an apical (in green) and an odd number of basal spins (dark red) in the contiguous Kagome plane. Closed loops within Kagome planes (not shown) are also possible, but they are not excitations. \textit{c-} A first string for $\mathbf{B}\parallel[110]$ runs in an average direction opposing $\mathbf{B}$. Its minimal block is made of an inverted spin in $\alpha-$chains (green) followed by spins on a $\beta-$chain crossing the same tetrahedron (dark red).}
\label{fig:UC_gen}
\end{figure}
%%%%%%%%%%%%%%%%%%%%%%%%%%%%%%

In the following, we use Monte Carlo simulations to characterize this sequence of transitions and analyze the scaling behavior of the magnetization and specific heat near each step. We then derive the corresponding critical fields theoretically, showing that their positions follow directly from the entropy–energy balance of a semi-infinite system. The combination of numerical and analytical approaches provides a consistent picture of how finite geometry can stabilize a cascade of Kasteleyn-like transitions in spin ice.

%%%%%%%%%%%%%%%%%%%    
\begin{figure}[bt]
\centering
\includegraphics[width=\linewidth]{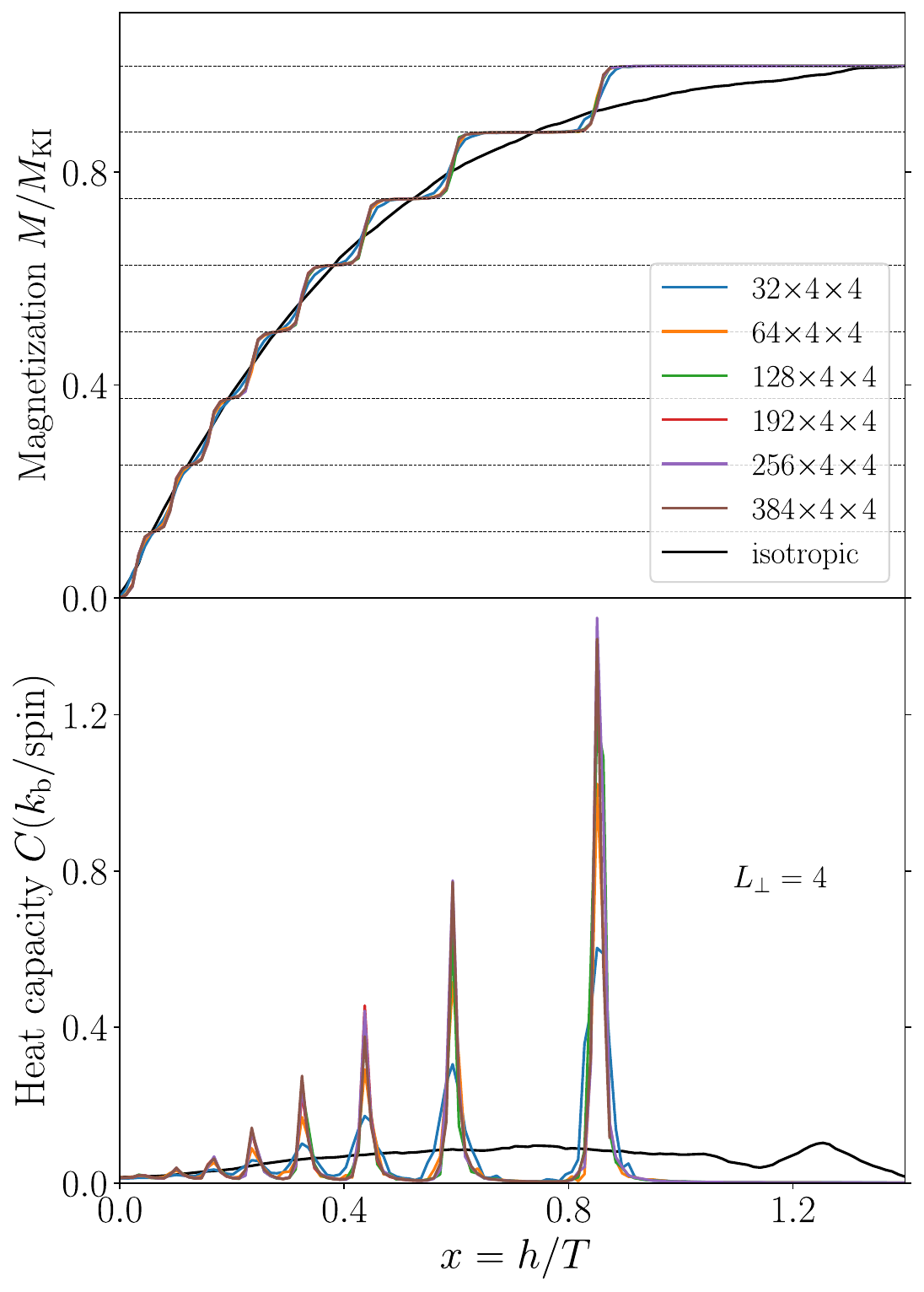}
\caption{Upper panel: Magnetization as a function of $x=h/T$ for lattices of different geometries with $\mathbf{B}\parallel[111]$ and $T/J_{\rm eff} = 0.3$.  The smooth black curve corresponds to the isotropic  case, $L\times L \times L$ with $L=8$. The stepped curves were measured in samples elongated along the [111] axis with dimensions $L_\parallel \times L_\perp \times L_\perp$, where $L_\perp = 4$ and $L_\parallel=8L_\perp \times n$, with the value of $n$ indicated in the legend.  The case $n=1$ contains the same number of spins as the isotropic system. Horizontal dashed lines mark the magnetization steps corresponding to the successive introduction of strings of negative magnetization that extend across the sample. Lower panel: Specific heat as a function of $x=h/T$. The height of the peak increases with $L_\parallel$. }
\label{fig:M.v.x.L4}
\end{figure}
%%%%%%%%%%%%%%%%%%%%%%%%%%%%%%

\section{Model and Background}

We consider the nearest-neighbors spin-ice Hamiltonian \cite{moessner1998relief,melko2004monte} consisting of classical Ising spins sitting at the corners of tetrahedra forming a pyrochlore lattice.  If one writes $\mathbf{S}_i=\sigma_i\hat{\mathbf{d}}_{i}$, where $\sigma_i=\pm 1$ and the $\hat{\mathbf{d}_i}$ point in the local $\langle 111 \rangle$ directions, one can write the spin-ice Hamiltonian~\cite{Harris1997} under a magnetic field $\mathbf{B}$ as, 
\begin{equation}
\mathcal{H} = J_{\rm eff} \sum_{\langle i j \rangle} 
\sigma_i \sigma_j 
- \sum_i \mathbf{h} \cdot \mathbf{S}_{i},
\tag{2.2}
\end{equation}
where $J_{\text{eff}}\equiv 1$ is antiferromagnetic and $h\equiv \mu B$, with $\mu$ the magnetic moment.

The ground states of this Hamiltonian at $h = 0$ are those spin configurations satisfying $\sum_{i=1}^4 \sigma_i = 0$ on every  tetrahedron, corresponding to the ``two-in, two-out'' ice rule (see Fig.~\ref{fig:UC_gen}). In a coarse-grained description, this local constraint enforces a divergence-free condition on the magnetization field. These states are macroscopically degenerate and give rise to a residual ground-state entropy, in agreement with Pauling’s estimate for proton disorder in water ice.  The lowest-energy excitation of the model is a single spin flip, which produces a pair of adjacent tetrahedra violating the ice rule: one with three spins pointing in and one out, and the other with the opposite configuration. In the coarse-grained picture, these defects correspond to positive and negative emergent magnetic monopole charges, respectively \cite{Castelnovo2008}. Flipping a different majoritarian spin belonging to one of the excited tetrahedra
restores the local two-in/two-out constraint while transferring the monopole to a neighboring tetrahedron that shares the flipped spin. This process can be repeated, allowing the
excitation to propagate through the lattice without additional energy cost; the \textit{string} of inverted spins that links the monopoles turns into a \textit{loop} when the monopoles meet and annihilate on a single tetrahedron. The presence of an applied magnetic field modifies this behavior: the associated
Zeeman energy assigns a finite magnetic tension to the string of reversed spins connecting the monopole pair \cite{szabo2025hidden}.

The behavior of this model under a magnetic field applied along [111] has been extensively investigated in Refs.~\onlinecite{moessner2003theory,isakov2004magnetization}. In this geometry, the field partially lifts the degeneracy of the spin-ice manifold and produces a sequence of distinct regimes.  The pyrochlore lattice may be regarded as a stacking of alternating Kagome and triangular planes along the [111] direction; in the latter, each spin is located at a corner shared by an up- and a down-pointing tetrahedron (see Fig.~\ref{fig:UC_gen}). Along this paper, when we mention a ``Kagome plane'' we will be talking about those Kagome planes perpendicular to [111].

If $T/J_{\rm eff} \ll 1$ the energy scale is set by the field, so that physical quantities are parametrized by a single thermodynamic variable $x \equiv h/T$. Also, the divergenceless condition guarantees that the spin dynamics is realized either through the inversion of spins in closed loops, or in extended \textit{strings} of spins spanning the whole system.

At intermediate field strengths, the system enters the Kagome-ice state (Fig.~\ref{fig:UC_gen}a), characterized by a magnetization plateau at the value $M_{\text{KI}}=1/3$ per magnetic moment. In this regime, spins in the Kagome planes remain frustrated, while those in the triangular planes are fully polarized. Within each Kagome plane perpendicular to the $[111]$ direction, closed loops of alternating
in--out spins can be flipped. Owing to their symmetry and alternating nature, such loops do not change the net magnetization along $[111]$ and therefore cost no energy within the nearest-neighbor model, even in the presence of a magnetic field. Moreover, their confinement to the Kagome planes preserves the topological sector of the system
\cite{jaubert2013topological}. 
Changes in the magnetization $M$ along this direction are tied to the creation or annihilation of \textit{strings} of inverted spins, spanning the system. They are filamentary in nature (see Fig.~\ref{fig:UC_gen}b), and travel, on average, along the $[111]$ direction; they close upon themselves through the periodic boundary conditions in such a way that they punch \textit{any} triangular plane perpendicular to [111] the same number of times; in other words, through the systematic inversion of apical spins, their creation imply a change of topological sector. As discussed in Ref.~\onlinecite{moessner2003theory}, this does not imply a phase transition as these excitations are analogous to defects in a paramagnet. Meandering of the string within a Kagome plane -- which takes place between two inverted apical spins in consecutive triangular planes-- incurs only a microscopic energy cost, owing to the alternating in--out nature of the meander. By contrast, the associated entropy gain scales logarithmically with the area of the Kagome plane. In an infinite system, these string excitations are therefore always thermodynamically favorable at finite temperature. As a consequence, they preclude any true phase transition between the spin-ice and Kagome-ice phases and instead introduce fluctuations that obscure the two-dimensional Kasteleyn transition in a perpendicular magnetic field \cite{moessner2003theory,isakov2004magnetization}. As the field is lowered, the system's three dimensional behavior is restored through the proliferation of these string objects.

Let us now consider an anisotropic sample with dimensions  $ L_{\parallel} \times L_\perp \times L_\perp$, where the longitudinal direction $L_\parallel$ runs parallel to the field and to [111]. In the limit $L_\parallel \to \infty$, with $L_\perp$ kept finite, the entropy of any meandering string segment inside a Kagome plane will be finite; we thus predict that the system would have a topological transition at the value of $x$ corresponding to the entry of the first string.  Furthermore, because the plane area is finite, the introduction of each additional string would entail a progressively smaller entropic change while incurring an identical energy cost. As a result, the balance between entropic gain and energetic cost depends on the number of strings already present, and the critical condition for entry will be no longer unique or continuous. Instead of a single transition, the system exhibits a sequence of discrete transitions, each associated with the entry of an individual string. This behavior is analogous to the sequence of transitions observed in anisotropic dimer models and two-dimensional spin-ice~\cite{bhattacharjee1983critical,ferreyra2018breakdown}. 

%%%%%%%%%%%%%%%%%%%    
\begin{figure}[bt]
\centering
\includegraphics[width=\linewidth]{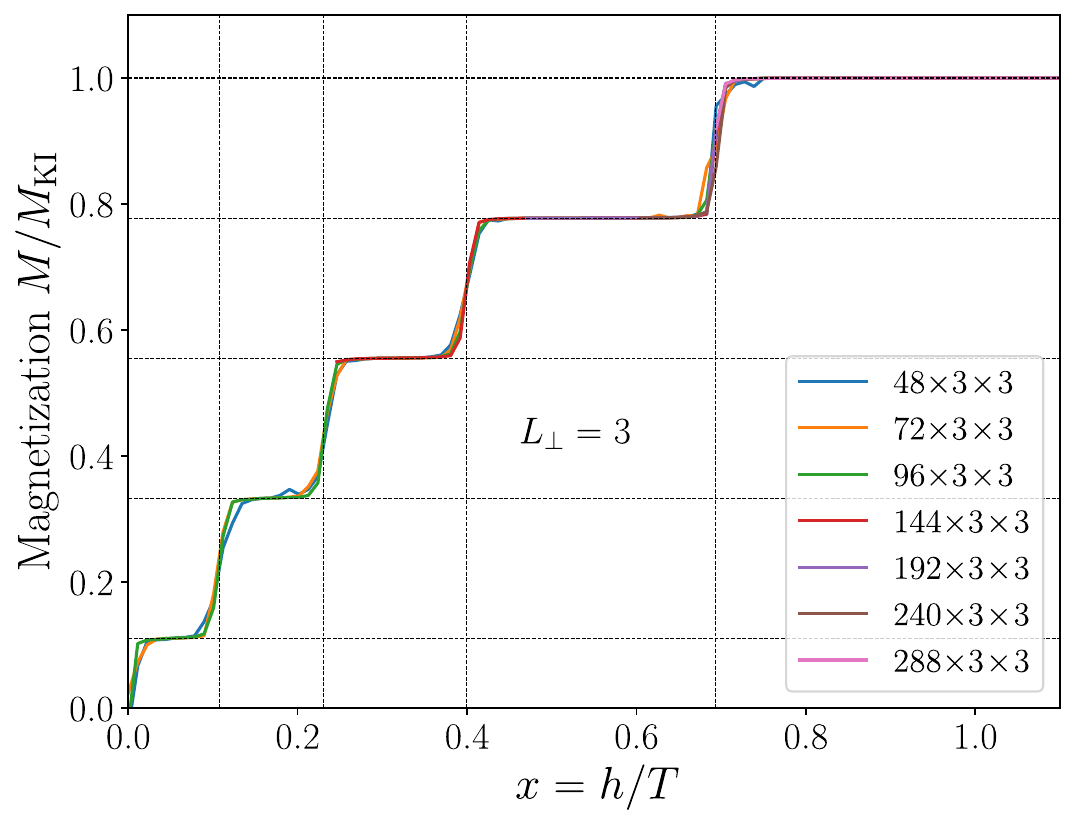}
\caption{Magnetization vs $x=h/T$ simulated for a system with $L_\perp=3$, and $L_\parallel$ in the direction of $\mathbf{B}\parallel[111]$. The vertical lines indicate the critical $x_i$ deduced from the exhaustive counting method (see text), while the equidistant horizontal ones indicate multiples of the magnetization of a string of spins running along [111]. Surprisingly, there is a half jump around $x=0$, with a diverging susceptibility in a symmetry point. Although this resembles a ferromagnetic critical point, the transition is first order but with no domains. %The horizontal dashed lines mark the magnetization steps in terms of single strings, while the vertical lines correspond to the values of $x_i$ calculated in the text. 
}
\label{fig:M.v.x.L3}
\end{figure}
%%%%%%%%%%%%%%%%%%%%%%%%%%%%%%

\section{Results}

We have performed Monte Carlo simulations of the nearest-neighbor spin-ice model using two different algorithms: one based exclusively on the inversion of closed spin loops (see \cite{szabo2025hidden}), and one employing the conventional single-spin-flip Metropolis algorithm; the approaches yield mutually consistent results. The data shown in the figures in this work were obtained using Metropolis. 

The simulations were carried out using a unit cell containing four spins forming a single tetrahedron, with rhombohedral geometry. The unit cell has three of its faces coincident with faces of the
tetrahedron (see Fig.~\ref{fig:UC_gen}a), such that periodic boundary conditions (PBC) close within each Kagome plane perpendicular to the field, as well as along the
$[110]$ direction. While boundary conditions are irrelevant in the thermodynamic limit, we will see that these PBC do produce observable effects for fields applied along $[110]$ (see Fig.~\ref{fig:110_4x1}), which are discussed in Section~\ref{sec:110}.

The data shown here were obtained at fixed temperature,
$T/J_{\rm eff} = 0.3$, and recorded as a function of the applied magnetic field at constant \(T\). This temperature is sufficiently high to ensure fast dynamics in the nearest-neighbor model, while still allowing the ratio $x = h/T$ to remain the sole relevant thermodynamic control parameter for the studied systems.

%%%%%%%%%%%%%

\subsection{$\mathbf{B}\parallel \mathrm{[111]}$}

The upper panel of Fig.~\ref{fig:M.v.x.L4} shows the magnetization 
{$M/M_{KI}$ along the [111] direction at $T/J_{\rm eff}=0.3$ as a function of $x=h/T$ for different sample geometries. The black curve was measured in an isotropic system of size $L \times L \times L$ with $L=8$ and displays a smooth increase from $M=0$ toward the Kagome-ice plateau at $M_{\text{KI}}$. In contrast, the stepped curves correspond to anisotropic samples of dimensions $L_\parallel \times L_\perp \times L_\perp$, with $L_\perp=4$ and $L_\parallel=8L_\perp \times n$, where $n$ ranges from 1 to 12. The case $n=1$ (blue) has the same number of spins as the isotropic sample. 

One can observe that, in the limit $L_{\parallel} \to \infty$, the magnetization approaches a sequence of strictly horizontal plateaux, accompanied by a vanishing specific heat $C$ (lower panel). In this system, which is governed by a single thermodynamic control parameter $x$, energy and magnetization play closely analogous roles. This makes clear that, upon moving away from the isotropic sample, the ability of $M$ to fluctuate is progressively suppressed over most values of $x$. Moreover, since changes in the magnetization along $[111]$ are mediated by system-spanning strings, each plateau can be identified with a distinct topological sector characterized by a fixed number of strings. The series of magnetization steps appearing at positions $x_i$, which sharpen systematically with increasing $n$, can be then understood as the successive entry of individual string excitations. 
%Each string changes the magnetization per spin by $\Delta m/m_{\text{KI}} = -2/L_\perp^2$
Each string changes the magnetization by $\Delta m_{\rm string}\equiv\Delta M/M_{\text{KI}} = -2/L_\perp^2$, 
corresponding to the spacing between the horizontal dashed lines in the figure. The lower panel shows the specific heat for the same simulations, with pronounced peaks at each $x_i$. %The peak heights scale linearly with $L_\parallel$, consistent with first-order transitions.
For large values of $L_\parallel$ the equilibration time becomes extremely long rendering MC equilibration practically unfeasible. As we will show below, the values of $x_i$ can be obtained by  counting the number of spin configurations of a finite-area Kagome plane pierced by a fixed number of strings.

There are several additional noteworthy features. We have seen that fluctuations in the magnetization and in the topological sector occur only at a discrete set of special values $x_i$. For this system, with even $L_{\perp}$, the thermodynamic state at $h = 0$ and $M = 0$ corresponds to an equal number of strings oriented along and against the $[111]$ direction. In the limit $L_{\parallel} \to \infty$, not only does the magnetization vanish, but the zero field magnetic susceptibility also tends to zero. At low fields, the system cannot increase its magnetization continuously. The same topological constraints that suppress fluctuations forbid incremental changes in $M$:
the only available mechanism is the replacement of an entire string opposing the field by one fully aligned with it. This process entails an entropy loss that is independent of the field and scales as \(\log L_{\parallel}\). Consequently, such a transformation cannot occur until this entropic penalty is compensated by a sufficiently large Zeeman energy gain, that is, only above a finite threshold field.  The resulting behavior, a low-susceptibility plateau followed by a sharp jump in magnetization, is reminiscent of a metamagnetic transition in an antiferromagnet, with entropy playing the role of the exchange energy cost. This analogy echoes the notion of a “topological metamagnet” introduced in Ref.~\cite{pili2022topological}.  Additionally, it is worth noting that, since the system is governed by a single thermodynamic variable, $x = h/T$, both the specific heat and the susceptibility exhibit the same underlying singular behavior.

Figure~\ref{fig:M.v.x.L3} shows the magnetization as a function of $x$ at the same fixed temperature, now for a lattice with an odd transverse size, $L_{\perp} = 3$. The curves are qualitatively similar to those in Fig.~\ref{fig:M.v.x.L4}, but display larger --and
consequently fewer-- steps. However, the behavior at $x = 0$ now contrasts sharply with that discussed in the previous paragraph for even $L_{\perp}$: while $M(h = 0) = 0$ remains unchanged, there are clear signatures of a divergent susceptibility, as expected for a symmetry-breaking ferromagnetic transition. Evidently, the system is now able to fluctuate, and it does so on a macroscopic scale.

This apparent discrepancy is resolved by noting that an odd number of apical spins in any plane perpendicular to the field necessarily implies an odd number of system-spanning strings. For a total of nine strings, the condition of vanishing magnetization at $h = 0$ can only be satisfied by having four strings polarized along each direction, together with a \emph{fluctuating string} that alternates its orientation. This fluctuating string continuously changes the imbalance between oppositely polarized strings and,
consequently, the magnetization. Such unrestricted fluctuations of $M$ account for the divergent susceptibility observed in the limit $L_{\parallel} \to \infty$. This behavior can also be interpreted as a Kasteleyn-like transition occurring precisely at $x = 0$, connecting a state with five strings polarized down and four up to its symmetric counterpart with five up and four down. The half-step observed in the magnetization (with the complementary half expected as $x \to 0^{-}$) provides further confirmation of this interpretation.

Having established the qualitative behavior, we next show how the sequence of transition points $x_i$ can be obtained analytically, with the help of some numerics.

\subsubsection*{Calculation of $x_{i}$ for $\mathbf{B} \parallel \mathrm{[111]}$}

In order to determine the critical value of $x_i=h_i/T$ for the introduction of $i$ strings, we begin by considering the polarized $\text{KI}$ state ($i=0$), and the emergence of strings, one by one, when the field is lowered or the temperature raised. We view the formation of a string as a process consisting of $L_\parallel$ successive steps. In order to construct the first string ($i=1$) we have to iterate $L_\parallel$ times through the following stages: 

\par $1-$ \emph{Flipping an apical spin.}
This marks the point at which the string enters an up-pointing tetrahedron and gains access to a given Kagome plane. Prior to the flip, the apical spin is fully aligned with the external magnetic field, resulting in an energy change
$\delta u_{\text{triang}} = +2h$. No entropic contribution is associated with this step ($\delta s_{\rm triang}=0$) since the entry point is uniquely determined by the exit tetrahedron of the previous step, and each tetrahedron contains a single apical spin.

\par $2-$ \emph{Flipping an odd number of alternating in--out basal spins.}
An odd number of basal spins must be flipped in the subsequent Kagome plane in order to preserve the spin-ice rule. This allows the string of inverted spins to exit the plane through a down-pointing tetrahedron. Each basal spin contributes only one third of its magnetic moment along the $[111]$ direction, with the sign determined by whether the spin points into or out of an up tetrahedron. As a consequence, consecutive pairs of spins cancel their Zeeman energy contributions, yielding a total energy change ${\delta u}_{\text{basals}} = +\tfrac{2}{3}h$, independent of the length of the flipped chain. The entropic contribution to the free-energy change associated with string creation, ${\delta s}_{\text{basals}}$, therefore arises solely from the number of distinct configurations available within a Kagome plane.

Within a mean-field treatment, the critical parameter $x_1$ for the entry of the first string can be obtained by equating to zero the free-energy change per unit length, $\Delta F_1/L_\parallel = \Delta f_1=\Delta u_1 - T\Delta s_1=0$, with the total change in energy and entropy calculated by repeatedly summing over the previous steps along the whole string. This gives
\begin{equation}
    x_1 \propto \frac{3}{8} \Delta s_1.
\end{equation}

The entropic change upon the creation of a string for narrow samples is not as direct to calculate as the energetic one. However, its dependence on the lateral area $A = L_\perp \times L_\perp$ can be estimated for large perpendicular areas. In this case, the initial, string-free entropy is proportional to the Kagome-ice entropy, $s_0 \propto L_\perp^2 s_{\text{KI}}$, with $s_{\text{KI}} \approx  0.0808$. The final entropy, $s_1$, retains most of this contribution, while adding a term that scales as $\log(A)$ associated with the uncertainty in the string’s exit point through any down tetrahedron (see Ref.~\cite{isakov2004magnetization} for a more detailed argument). Hence, $\Delta s_1 \propto \log(A)$ and $x_1 \propto \frac{3}{8} \log(A) \to \infty$
in the limit $L_\perp \to \infty$. A similar calculation was previously carried out in Refs.~\cite{moessner2003theory,isakov2004magnetization}, with the important corollary that in the thermodynamic limit the Kagome-ice state is not a distinct phase, but rather a smooth continuation of the spin-ice manifold.
We will show now that if the thermodynamic limit is taken while keeping the transverse area $A = L_\perp^2$ finite, the situation changes qualitatively: the Kagome-ice state becomes a distinct phase, and the same mechanism gives rise to a cascade of transitions occurring at a discrete set of $x_i$.

The above discussion was elaborated around the creation of the first string of inverted spins, but it will be valid for any other value of $i$. The Zeeman energy change is proportional to the change in magnetisation when a string is added and, thus, independent of the number of strings already present in the system:  $\Delta u_n = 8/3h$. The mean-field general expression for the critical value of $x_i$ for the passage between $(i-1) \to i$ strings is thus
\begin{equation}
    x_i = \frac{3}{8} (s_i - s_{i-1}) \, ,
    \label{eq:x_n}
\end{equation}
with the $s_i$ the entropy of a Kagome plane crossed by $i$ strings.
Given their finite size, boundary conditions are now important and the value for the Kagome-ice entropy cannot be used as a general result. Instead, one has to calculate for each system the number of configurations $\Omega_i$ for $i$ strings entering a given Kagome plane through a set of known up pointing tetrahedra. $\Omega_i$ is not unique and would generally depend on the relative location of the entering points for different strings.  We calculated these $\Omega_i$ by exhaustively visiting all possible Kagome plane configurations, and counting the exact number of those having $i$ strings entering the plane through fixed positions. Fig.~\ref{fig:omega} gives an example for the values of $\Omega_0$, $\Omega_1$ and $\Omega_2$ for $L_\perp=2$ and $L_\perp=3$.  The very nature of the approach constraint us to very small Kagome planes.  From these $\Omega_i$ we calculate the average plane entropy $\overline{s_i}$, resulting from strings crossing each plane through $i$ up tetrahedra.

%%%%%%%%%%%%%%%%%%%    
\begin{figure}[bt]
\centering
\vspace{2ex}
\includegraphics[width=\linewidth]{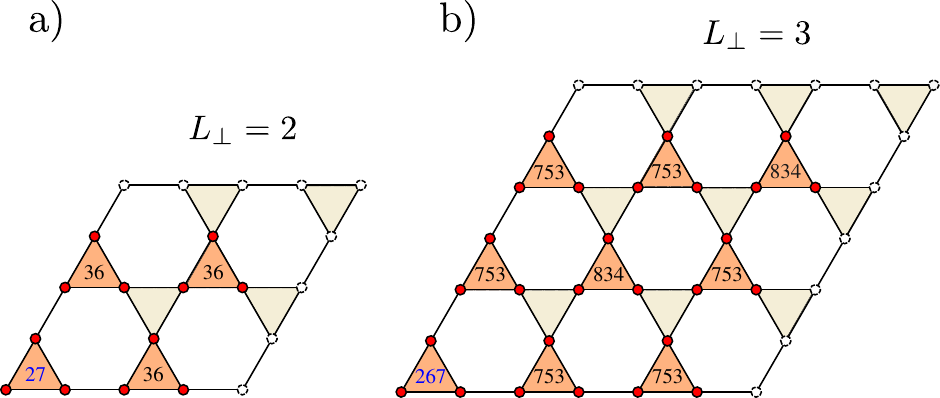}
\caption{Number of spin configurations $\Omega_1$ (in blue) with one string going through a fixed up tetrahedron and out through any down tetrahedron (painted gray). Along the text we take the fixed up tetrahedron as the $(1,1)$ one. $\Omega_2(i,j)$ (in black) designates this number for two strings, one passing through the (1,1) and the other through one of the different up tetrahedra at the location $(i,j)\neq (1,1)$. For 0 strings, $\Omega_0=9$ for $L_\perp=2$, and $\Omega_0=42$ for $L_\perp=3$. $\Omega_n$ was calculated through the exhaustion method up to lattices with 9 unit cells in the direction perpendicular to the [111] plane.} \label{fig:omega}
\end{figure}
%%%%%%%%%%%%%%%%%%%%%%%%%%%%%%

For $L_\perp = 2$, the possible configurations can be enumerated directly. There are $\Omega_0 = 9$ configurations with no strings, and $\Omega_1 = 27$ configurations with a single string entering through the $(1,1)$ up tetrahedron (marked in blue in Fig.~\ref{fig:omega}a)) or through any other one. There are $\Omega_2 = 36$ configurations with two strings, where one enters through the $(1,1)$ up tetrahedron and the other through one of the remaining $(i,j)$ up tetrahedra -- these cases being equivalent by symmetry marked in black in Fig.~\ref{fig:omega}a)). As required by symmetry, $\Omega_3 = 27 = \Omega_1$; since $\Omega_3 < \Omega_2$, only two strings (and therefore a maximum of two topological transitions) can enter the system for any field $\mathbf{B}$ oriented along [111]. Equivalently, introducing a third string would correspond to a negative magnetization in a positive field, which is energetically forbidden.
Using these degeneracies, we obtain two transitions located at $x_1 = \tfrac{3}{8}\log(3) \approx 0.412$ and $x_2 = \tfrac{3}{8}\log(4/3) \approx 0.108$, in good agreement with the values extracted from the Monte Carlo simulations (not shown).

The case $L_\perp=3$ introduces the possibility of fluctuations in the value of $\Omega_i$ for each plane, depending on the strings entering points. Using the exhaustion counting, we obtain $\Omega_0=42$, and $\Omega_1=267$ (marked in blue in Fig.~\ref{fig:omega}b)). This means a first string entering at $x_1=3/8\times \log(267/42)=0.6936$. The number of configurations for two strings, one through $(1,1)$ and a second through $(i,j)$, are marked in black in Fig.~\ref{fig:omega}b). Here, as in the general case, the precise number of configurations depends on the relative location of the two entry points, $\Omega_2=\Omega_2 (i,j)$, with $\Delta \Omega_2/\overline{\Omega_2} \approx 0.05$. We calculate the average value of $s_2$ to be $\overline{s_2}=6.6496$. Using this, we obtain $x_2 \approx 0.399$. For three strings, the fluctuations in $\Omega_3$ are approximately twice as big as for two. There are a total of $120411$ configurations with three strings, and there are $9!/(6!3!)=84$ possible ways of passing choosing the entering points, giving an average of $\overline{\Omega_3}=1433.3$ configurations with $3$ strings going through fixed up tetrahedra. This gives (assuming that the distribution of $\Omega_3$ is narrow enough) $x_3 \approx 0.231$. For $4$ strings, $x_4 \approx 0.1076$.
The estimated values for $x_i$ are plotted as vertical lines in Fig. \ref{fig:M.v.x.L3}; we observe a very good coincidence with the jumps in the Monte Carlo simulations. 

We have seen that for this field orientation and due to the local, divergenceless constraint, magnetization changes can only happen through the introduction or removal of strings of negative magnetization.  In a case such as $L_\perp=4$, with 16 up-tetrahedra per Kagome %(111)
planes, and thus $16$ potential strings of inverted spins, the maximal entropy state ($M=0$) would be reached with half of the up tetrahedra punched by strings of inverted spins.  As we discussed before, this is seen in Fig. \ref{fig:M.v.x.L4}, where the magnetization goes from $M_{\text KI}$ to zero through $8$ equal steps with a size given by the magnetization of a single string $\Delta m_{\rm string}$ (vertical dotted lines).  If we apply the same analysis to  $L_\perp$ odd, like $L_\perp=3$, we would conclude that there must be a fractional number of strings, $9/2$ in this case, for $M=0$ at $x=0$. Indeed, the Monte Carlo simulations show, as $x$ is decreased towards $0$, four transitions with a full jump $\Delta m_{\rm string}$ (vertical dotted lines).  The ``half-jump'' observed at $x=0$ can be readily understood by noticing that the other half of the jump at finite $L_\parallel$ is at infinitesimally small \textit{negative} fields, completing a whole magnetization jump. In other words: at the point $M=0$ at $x=0$ fluctuations take the magnetization from negative to positive with a jump totaling $\Delta m_{\rm string}$. As with spontaneous symmetry breaking in a ferromagnet, the response to an infinitesimal field diverges for $L_\parallel \to \infty$. Unlike a soft ferromagnet, no domains or domain wall movement are implied in this giant susceptibility, only the inversion of a single string of spins. We can think of this as a \textit{topological ferromagnetic transition}, corresponding to the limit of the topological metamagnetic transition~\cite{pili2022topological} with $h \to 0$. This property will be present for any system with odd $L_\perp$.

\subsection{$\mathbf{B} \parallel \mathrm{[110]}$} \label{sec:110}

%%%%%%%%%%%%%%%%%%%    
\begin{figure}[bt]
\centering
\includegraphics[width=\linewidth]{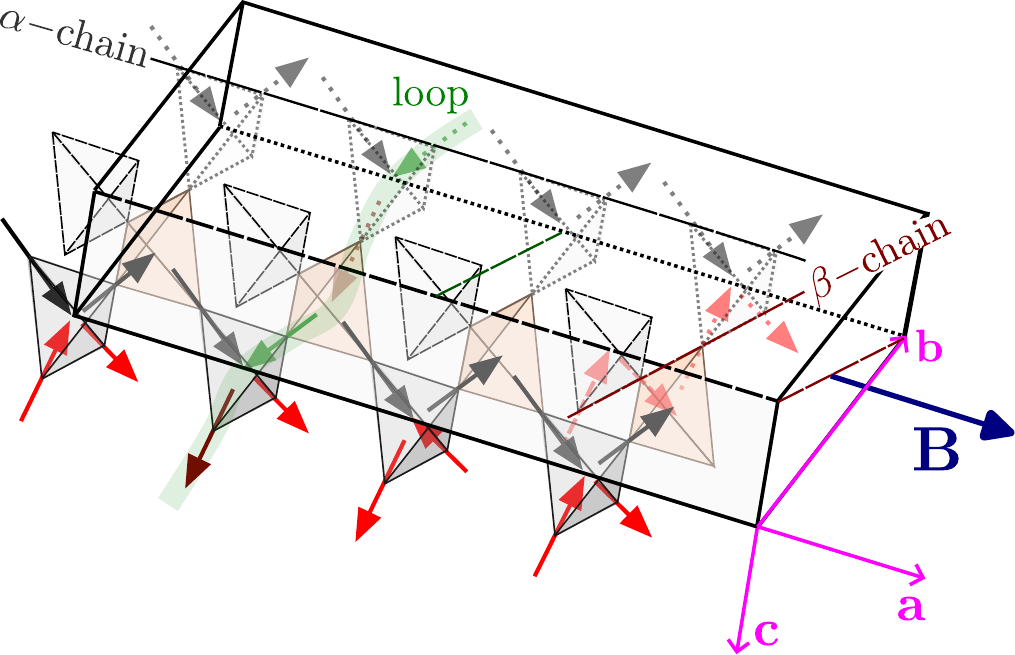}
\caption{A sample of $4\times 1\times 1$ rombohedral unit cells used in our simulations, with non-perpendicular cell-vectors $\mathbf{a}$, $\mathbf{b}$, and $\mathbf{c}$ of equal modulus. We mark up (orange) and down (grey) tetrahedra, $\alpha-$chains parallel to the field $\mathbf{B} \parallel \mathbf{a} \parallel [110] $, and perpendicular $\beta-$chains. The real set of spins and tetrahedra are drawn in continuous line, but we include copies of them linked by PBC in different types of line to help visualization. There are inverted $\alpha-$spins, although no string has entered the system: this PBC allow the insertion of spin \textit{loops} (made of inverted $\alpha$, in green, and $\beta-$ spins, in dark red) contained into Kagome planes. These loops change de magnetization along [110], but do not change the topological sector nor break the ice rules. They are responsible for the lack of saturation and the slight slope in the diverse plateaux observed in the magnetization curves (see Fig.~\ref{fig:110_MrhoC}).}
\label{fig:110_UC_scheme}
\end{figure}

We now turn to the possibility of observing a cascade of Kasteleyn-like transitions in geometrically confined samples when the field is aligned along the [110] axis. What made Kasteleyn-like transition possible for $\mathbf{B} \parallel[100]$ for bulk samples was the inability of the filaments to move perpendicular to the field direction, while what made them seem impossible for bulk $\mathbf{B}\parallel[111]$ was the possibility of increasing the entropy without bounds by their meandering in \textit{planes} perpendicular to the field. In the previous section, we showed that a Kasteleyn-like transition is possible by constraining the size of these planes. We will now see that the $\mathbf{B}\parallel[110]$ case is analogous,  with the difference that the field direction now splits the spin system in chains rather than planes (see Fig.~\ref{fig:110_UC_scheme}).
Although for bulk samples there is \textit{no} Kasteleyn transition at any finite value of $x$ for $\mathbf{B}\parallel[110]$~\cite{fukazawa2002magnetic,benton2022spin}, we will recover a transition (or, rather, a cascade of transitions) by restricting the length of $\beta-$chains (see Fig. \ref{fig:UC_gen}a). 

%%%%%%%%%%%%%%%%%%%    
\begin{figure}[bt]
\centering
\vspace{2ex}
\includegraphics[width=\linewidth]{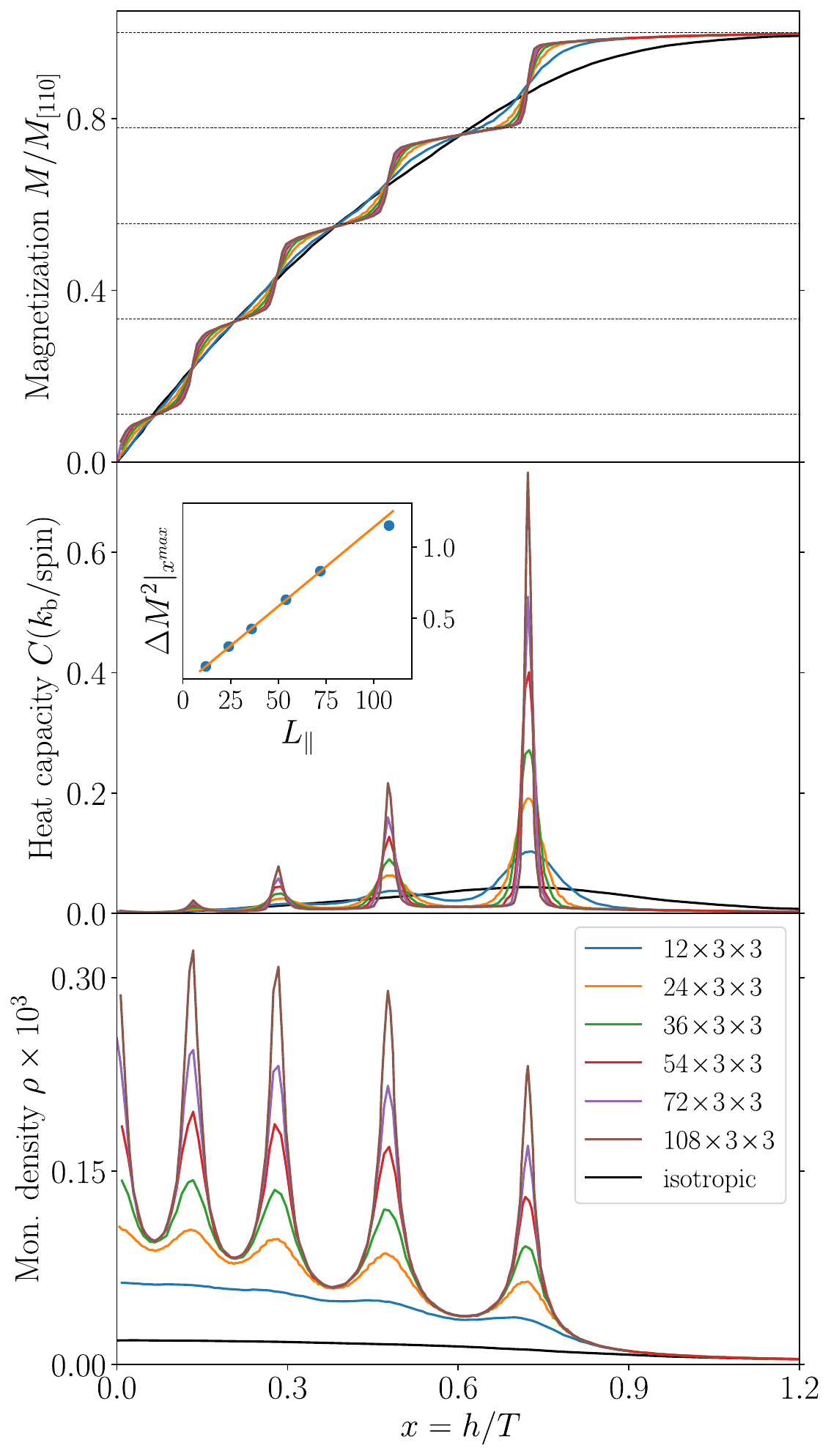}
\caption{Magnetization (upper panel), specific heat $C$ (middle panel), and number density of monopoles per tetrahedron $\rho$ (lower panel) for samples of size $L_\parallel\times L_\perp \times L_\perp$ with $\mathbf{B} \parallel [110]$, with the longest side of the sample oriented along the field. 
$L_\perp=3$ and $L_\parallel=4L_\perp \times n$, where $n$ ranges from 1 to 9.
The inset to the middle panel shows the increase in height of the peak in magnetic susceptibility $\chi$ as a function of $L_\parallel$ at $x_1$, suggesting a first order transition. The same behavior is observed in $C$. Remarkably, $\rho$ seems to copy some features from the response functions.} 
\label{fig:110_MrhoC}
\end{figure}
%%%%%%%%%%%%%%%%%%%%%%%%%%%%%%

Beyond providing a second example in which no transition occurs in the
three-dimensional thermodynamic limit, we will use this case to examine the
field dependence of the density of single monopoles and to investigate how the boundary conditions in the directions along which the sample remains finite may have an effect on the phase transitions.

The black curve on the top panel of Fig.~\ref{fig:110_MrhoC} shows the magnetization along the [110] direction $M/M_{[110]}$ vs $x$ for an isotropic sample 
at low temperature. As discussed in ref. \onlinecite{fukazawa2002magnetic} for macroscopic samples, this case is not unlike that of a paramagnet, with a saturation value per magnetic moment given by $M_{[110]}=(\sqrt{2/3}\times 2)/4$.  In each tetrahedron, two spins belong to the same $\alpha$ chain and are polarized by the applied field, with one spin pointing in and the other pointing out (see Fig.~\ref{fig:UC_gen}). The remaining two spins, belonging to a $\beta$ chain, are insensitive to the external field but remain subject to the internal constraints: to satisfy the spin-ice rule in the low-temperature regime $(T/J_{\rm eff} \ll 1)$, they must adopt an in--out or out--in configuration. For an isolated tetrahedron, this choice would represent a local degeneracy. However, since each spin --and in particular each $\beta$ spin-- is shared between two neighboring tetrahedra, the freedom is in fact global, resulting in a twofold degeneracy per $\beta$ chain. Consequently, unlike the previous field orientation, the residual entropy associated with this direction is subextensive.

%%%%%%%%%%%%%%%%%%%    
\begin{figure}[bt]
\centering
\vspace{2ex}
\includegraphics[width=\linewidth]{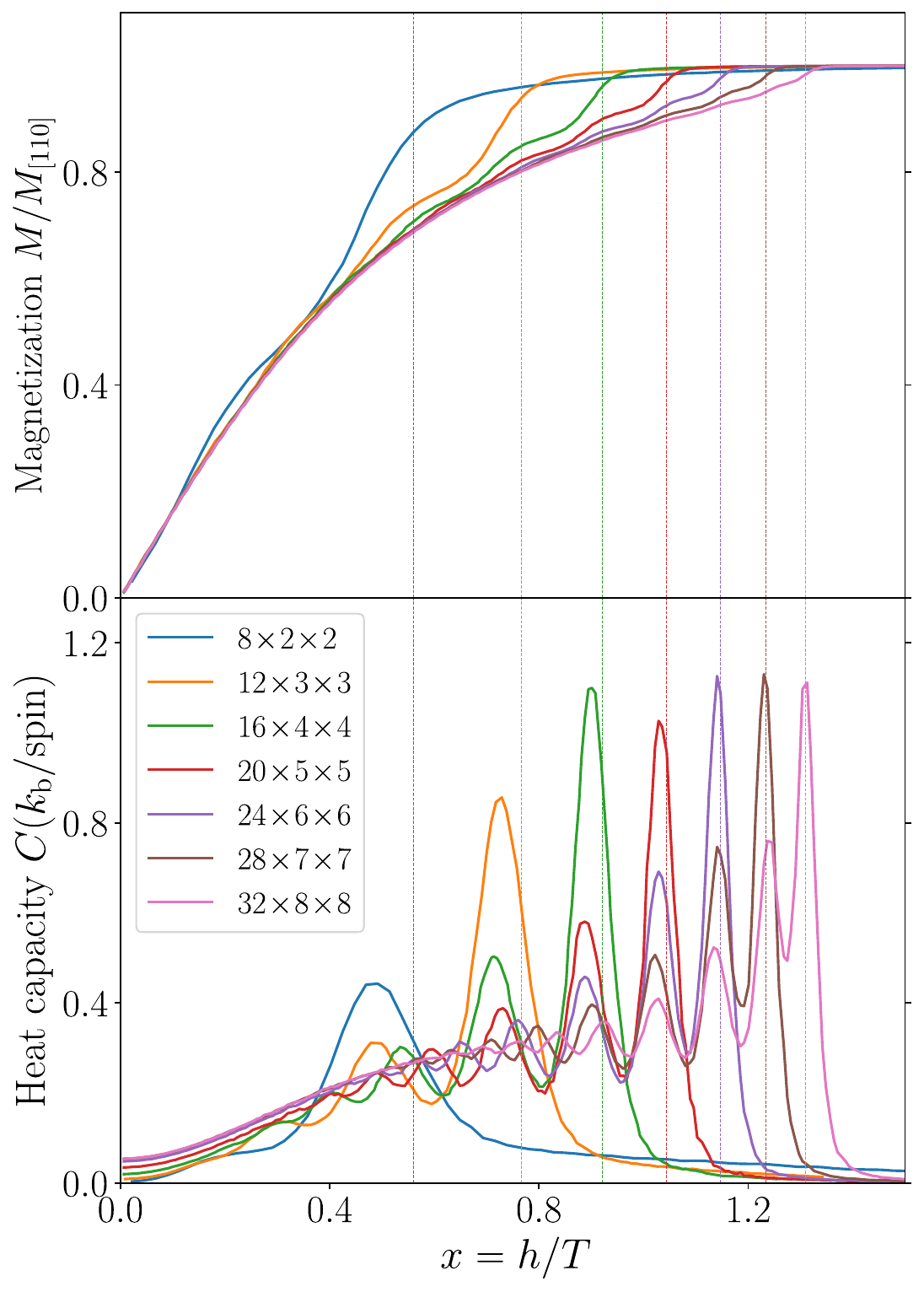}
\caption{Magnetization (top panel) and specific heat (lower panel) for samples of size $n(4\times 1 \times 1)$. The dashed, vertical lines, mark the position where the first string is expected to enter the system,
$x_{1}(L)=\sqrt6/4\times \log \left(L_\perp+\frac{1}{2}\right)$. The prediction works very well for big $n$ (i.e., big $L_\perp$), and it is able to predict the critical value for the entrance of further strings (for example, it can be seen that $x_1(L_\perp-1)\approx x_2(L)$ or, in general, $x_1(L_\perp-m)\approx x_m(L)$. However, for small lattice sizes the transitions move to smaller values of $x$, where the system scape magnetic saturation through the entrance of loops (see Fig.~\ref{fig:110_UC_scheme}). These non-topological excitations where not taken into account in the expression of $x_1$.}
\label{fig:110_4x1}
\end{figure}
%%%%%%%%%%%%%%%%%%%%%%%%%%%%%%

Despite the above considerations, there is a marked contrast between the smooth
magnetization curve observed for the isotropic sample and the behavior of the anisotropic systems, as was already the case for $\mathbf{B} \parallel [111]$. Here the magnetization also evolves through a sequence of plateaux -- albeit not perfectly flat -- separated by abrupt jumps. These jumps become increasingly sharp and well defined as $L_{\parallel}$ increases at fixed $L_{\perp} = 3$. The middle panel displays the corresponding behavior of the specific heat $C$. The peak heights scale linearly with $L_{\parallel}$, confirming the first-order character of the transitions. The inset shows this linear scaling for the susceptibility peak at $x_1$; an analogous scaling behavior is observed for $C$
\footnote{We note a slight deviation from linear scaling for the largest values of $L_{\parallel}$ considered. It may be attributed to single-spin excitations or/and to the presence of closed-loop excitations (discussed in the main text), which can cut system-spanning strings into segments shorter than $L_{\parallel}$ for large samples.}.

The lower panel of Fig.~\ref{fig:110_MrhoC} displays the density of single monopoles, $\rho$, as a function of $x$ for the different system sizes. For the isotropic sample, shown by the black curve, $\rho$ exhibits the expected smooth behavior: it starts from a baseline value of order \(10^{-4}\) monopoles per tetrahedron (its maximum possible value being one single monopole per tetrahedron) and decreases with increasing field as the system approaches saturation. In striking contrast, for the anisotropic samples the monopole density develops pronounced peaks at the same values $x_i$ at which anomalies appear in the response functions (middle panel). The height of these peaks grows approximately in proportion to the system length $L_{\parallel}$. We discuss this behavior below, as it provides insight into the mechanism of string formation and sheds light on additional aspects of these unusual
transitions.

Our working regime ensures a baseline of extremely low density of monopoles. At $B = 0$, $\rho \approx 10^{-4}$ corresponds to an average of approximately $1/20$ monopole pairs present at any given time in a system of size $72 \times 3 \times 3$. More precisely, this means that a single monopole pair is present only for about one twentieth of the measurement time: a pair is sporadically created and subsequently annihilated at different locations in the sample.
At fields sufficiently strong to saturate the magnetization, the lifetime of such pairs is even shorter. In this regime, separating two opposite charges reduces the magnetization by an amount proportional to their separation projected along the field direction. The resulting Zeeman energy cost is positive and acts as a tension on the segment of reversed string connecting the monopoles, driving it to contract and causing the pair to annihilate shortly after its creation.  Near the critical value $x_1$, however, this string tension is compensated by an entropic force that tends to push the charges apart in an exploratory, diffusive motion. Although the charges ultimately annihilate, they are likely to do so only after winding once around the sample, having traveled a distance of order $L_{\parallel}$ and thereby creating a system-spanning string of inverted spins. Assuming a random-walk
dynamics at such compensation points \(x_1\) (and similarly at other critical values $x_i$), the expected lifetime of a monopole pair scales as $L_{\parallel}^2$. It then follows naturally that the average monopole density $\rho$ at these critical points increases in proportion to $L_{\parallel}$, consistent with the numerical observations.

\subsubsection*{Calculation of $x_{i}$ for $\mathbf{B} \parallel \mathrm{[110]}$}

Once again, in an infinite sample any change in magnetization along [110] is associated with the entry of strings of inverted spins that run, on average, along the direction of the applied field. Their construction is analogous to, and in fact somewhat simpler than, the case of a field along $[111]$; here we focus on the creation of the first string. The iterative two-stage process now proceeds along the $[110]$ direction, whose linear extent is $L_{\parallel}$. 
The first stage corresponds to the string entering an up-pointing tetrahedron through an $\alpha$ spin, contributing solely to the energy term. The second stage involves flipping $ 0, 1, 2, \ldots$ spins along the $\beta$ chain that crosses the same up-pointing tetrahedron. As before, this second stage is the one responsible for the entropic contribution; however, it carries no Zeeman energy cost and proceeds along a straight line, terminating when a second $\alpha$ spin is flipped further along the $[110]$ axis. 

Each tetrahedron contains two $\beta$ spins, and there is a single tetrahedron per unit cell (Fig.~\ref{fig:110_UC_scheme}); each $\beta-$chain consists of $2L_{\perp}$ spins. For a fixed entry to the $\beta-$chain and any exit point, there are $\Omega_1=2L_{\perp}+1$ possible configurations (including the option of flipping no spins and exiting through the second $\alpha-$ spin in the up tetrahedron that serves as entrance). The  corresponding entropy is therefore $\log(2L_{\perp}+1)$. In the pristine, saturated state the $\beta$ chains are doubly degenerate $\Omega_0=2$; consequently, the entropy change associated with the creation of the first string is
$\Delta s_1 = s_1 - s_0 = \log{(2L_{\perp}+1)/2},$ and the expected critical value is
\begin{equation}
    x_1(L_\perp) = \frac{\sqrt6}{4} \log \left ( L_\perp+\frac{1}{2}\right ) \, .
    \label{eq:H110-x_1}
\end{equation}

We have tested this prediction for the entry of the first string by performing simulations on lattices of size $n(4\times1\times1)$,
with $n=2,\ldots,8$. These samples preserve a fixed aspect ratio and are oriented such that the longest dimension lies along the $[110]$ direction of the applied magnetic field (see Fig.~\ref{fig:110_UC_scheme}). The magnetization and specific heat as functions of $x$ are shown in the upper and lower panels of Fig.~\ref{fig:110_4x1}, respectively. The colored vertical lines indicate the values $x_1(L_\perp)$ predicted by Eq.~\ref{eq:H110-x_1}.

While the prediction is rather poor for the smallest samples, it improves systematically as $n$ increases (i.e., as $L_\perp$ grows). For sufficiently large $n$, not only does $x_1(n)$ accurately predict the critical value of $x$ at which the first string enters a system of size $n$, but it also provides a good approximation to the critical value for the second string in a system of size $n+1$, the third string in a system of size $n+2$, and so on. For instance, the red vertical line gives a reasonable estimate for the first
string in a system with $n=5$, and is equally accurate for the second string in a system with $n=6$, the third string for $n=7$, etc.

We do not address here the physics of multiple strings for $\mathbf{B}\parallel[110]$. Instead, we explain why Eq.~\ref{eq:H110-x_1} fails for small system sizes and why its accuracy improves progressively with increasing $n$. As shown in Fig.~\ref{fig:110_UC_scheme}, the periodic boundary conditions employed in these simulations allow for the existence of closed loops. These loops satisfy the ice rules, involve $\alpha$ spins, and therefore carry a magnetic moment, but they are \textit{not} strings. Much like in a conventional paramagnet, such excitations provide a mechanism for reducing the magnetization without the creation of system-spanning strings.
This effect is already apparent in Fig.~\ref{fig:110_MrhoC}: in contrast to the case $\mathbf{B}\parallel[111]$, the magnetization plateaux exhibit a finite slope, and saturation is approached more gradually than in our previous study. Both features are clear signatures of magnetization fluctuations that do \textit{not} involve strings, but instead arise from these loop excitations. Their presence modifies the counting of configurations associated with a string, thereby explaining the breakdown of Eq.~\ref{eq:H110-x_1} for small system sizes.

For increasing $n$, however, wider samples admit the first string only at larger values of $B$. At the corresponding larger values of $x$, such paramagnetic loop excitations become increasingly rare, and the assumptions underlying the prediction for $x_1$ are better satisfied. Consistently, as shown in Fig.~\ref{fig:110_4x1}, the agreement with Eq.~\ref{eq:H110-x_1} improves (lower panel) as the magnetization approaches saturation (upper panel).

\section{Discussion and Conclusions}

We have investigated the magnetization process of the nearest-neighbor spin-ice model in a magnetic field applied along the global [111] and [110] directions, with particular emphasis on the role of the sample shape. While no Kasteleyn-type transition is present for these field orientations in the three dimensional thermodynamic limit, we have shown that restricting the transverse dimensions of the system qualitatively changes the physics. For samples elongated along the field direction and with finite transverse area, the divergence-free constraint quantizes the number of string excitations that can span the system, which can give rise to a cascade of discrete topological transitions.

For a magnetic field applied along the 
$[111]$  direction, these transitions manifest as sharp steps in the magnetization, accompanied by pronounced peaks in the specific heat and magnetic susceptibility. The case of a field applied along $[110]$ exhibits many similarities; however, the presence of loop excitations prevents the formation of perfectly flat magnetization plateaux, instead giving rise to a finite slope. Monte Carlo simulations show that the height of the specific heat and suceptibility peaks scale linearly with the %system length
longitudinal length of the system} consistent with first-order transitions between distinct topological sectors. An analytical treatment based on the entropy–energy balance of a semi-infinite system reproduces the sequence of critical fields and accounts for their dependence on the transverse system size.

Finite geometry therefore plays a constructive role: instead of rounding the magnetization process, it stabilizes a sequence of topological transitions that merge into a smooth crossover only in the isotropic limit. This demonstrates that topological order in spin ice can be controlled not only by field direction and temperature, but also by sample shape.

A particularly notable effect occurs for systems with an odd number of transverse unit cells. In this case, symmetry requires a half-integer number of system-spanning strings at zero field, leading to a macroscopic magnetization jump at $h=0$. In the limit of infinite system length, the associated susceptibility diverges, despite the absence of domain formation or conventional symmetry breaking. The magnetic response is instead governed by the inversion of a single topological string.

This behavior may be viewed as a topologically enhanced susceptibility, in which an infinitesimal magnetic field induces a macroscopic change in magnetization through a topological mechanism. Because the response is protected by global constraints rather than fine-tuned energetic competition, it is robust to microscopic perturbations. This suggests a potential route toward magnetic-field sensing based on geometrically constrained frustrated magnets.

\begin{acknowledgments}
We acknowledge the Agencia Nacional de Promoción Científica y Tecnológica (ANPCyT) Argentina, for awarding grants  PICT 2022-11-00046 and PICT 2022-11-00100, which, although financed by the Inter-American Development Bank (IDB), were later discontinued following changes in national funding policy.
\end{acknowledgments}

%%%%%%%%%%%%%%%%%%%%%%%%%%%%%%
\bibliographystyle{apsrev4-2}.
\bibliography{postdoc}% Produces the bibliography via BibTeX.

\end{document}